\documentclass[sigconf, 9pt]{acmart}
\newcommand{\fig}[1]{Fig.~\ref{#1}}

\newcommand{\shortpar}[1]{\textbf{#1}}

\newcommand\size{.16}

\usepackage{acronym}
\usepackage{subfig}
\usepackage[justification=centering]{caption}
\usepackage{graphicx}

\AtBeginDocument{%
  \providecommand\BibTeX{{%
    \normalfont B\kern-0.5em{\scshape i\kern-0.25em b}\kern-0.8em\TeX}}}

\setcopyright{acmcopyright}
\copyrightyear{2023}
\acmYear{2023}
\acmDOI{XXXXXXX.XXXXXXX}

\acmConference[mmWaveSys]{First ACM Workshop on mmWave Sensing Systems and Applications}{November 12, 2023}{Istanbul, Türkiye}
\acmPrice{15.00}
\acmISBN{978-1-4503-XXXX-X/18/06}

\acrodef{isac}[ISAC]{Integrated Sensing And Communication}
\acrodef{tx}[TX]{transmitter}
\acrodef{rx}[RX]{receiver}
\acrodef{trn}[TRN]{beam-training}
\acrodef{to}[TO]{timing offset}
\acrodef{fo}[CFO]{carrier frequency offset}
\acrodef{los}[LoS]{line-of-sight}
\acrodef{bp}[BP]{beam pattern}
\acrodef{cir}[CIR]{channel impulse response}
\acrodef{md}[$\mu$D]{micro-Doppler}
\acrodef{aod}[AoD]{angle of departure}
\acrodef{ekf}[EFK]{Extended Kalman Filter}
\acrodef{bp}[BP]{beam pattern}
\acrodef{snr}[SNR]{signal-to-noise ratio}
\acrodef{gt}[GT]{ground-truth}
\acrodef{mmwave}[mmWave]{millimeter-wave}
\acrodef{lo}[LO]{local oscillator}
\acresetall
\begin{document}

\title{An Experimental Prototype for Multistatic Asynchronous ISAC}

\author{Marco Canil}
\email{marco.canil@phd.unipd.it}
\orcid{0000-0001-8037-7497}
\affiliation{%
  \institution{University of Padova}
  \city{Padova}
  \country{Italy}
}

\author{Jacopo Pegoraro}
\email{jacopo.pegoraro@unipd.it}
\affiliation{%
  \institution{University of Padova}
  \city{Padova}
  \country{Italy}
}

\author{Jesus O. Lacruz}
\email{jesusomar.lacruz@imdea.org} 
\affiliation{
\institution{IMDEA Networks}
\city{Madrid}
\country{Spain}}

\author{Marco Mezzavilla}
\email{mezzavilla@nyu.edu} 
\affiliation{
\institution{New York University}
\city{New York}
\country{United States}}

\author{Michele Rossi}
\email{michele.rossi@unipd.it} 
\affiliation{
\institution{University of Padova}
\city{Padova}
\country{Italy}}

\author{Joerg Widmer}
\email{joerg.widmer@imdea.org} 
\affiliation{
\institution{IMDEA Networks}
\city{Madrid}
\country{Spain}}

\author{Sundeep Rangan}
\email{srangan@nyu.edu} 
\affiliation{
\institution{New York University}
\city{New York}
\country{United States}}


\begin{abstract}
    We prototype and validate a \textit{multistatic} \ac{mmwave} \ac{isac} system based on IEEE~802.11ay. Compensation of the clock asynchrony between each \ac{tx} and \ac{rx} pair is performed using the sole \ac{los} wireless signal propagation. As a result, our system provides concurrent target tracking and \ac{md} estimation from multiple points of view, paving the way for practical multistatic data fusion. Our results on human movement sensing, complemented with precise, quantitative \ac{gt} data, demonstrate the enhanced sensing capabilities of multistatic ISAC, due to the spatial diversity of the receiver nodes. 
\end{abstract}






\maketitle

\section{Introduction}\label{sec:intro}

\acf{isac} reuses wireless network devices for sensing the surroundings by applying the radar principles to communication signals.
The strength of this approach lies in the ubiquity of wireless communication devices which avoids the costly deployment of dedicated sensors and allows exploiting spatial diversity and boost sensing accuracy thanks to \emph{multistatic} settings \cite{han2023multistatic}.
However, communication devices are based on separated \ac{tx} and \ac{rx} nodes with \emph{asynchronous} \acp{lo}, which make conventional radar techniques unusable~\cite{zhang2022integration}.
Several methods exist to mitigate the impact of such asynchrony~\cite{zhang2022integration}, but none have been experimentally validated on multistatic scenarios.

We fill this gap by presenting the first experimental prototype of \textit{asynchronous and multistatic} ISAC in the \ac{mmwave} frequency band, based on the IEEE~802.11ay standard at 60~GHz. The LO asynchrony is compensated for by exploiting the \ac{los} propagation of the signal between each \ac{tx} and \ac{rx} pair. 
Our results on human movement sensing show that the system can accurately track people and estimate their \ac{md} signature from multiple points of view, providing enriched and more reliable movement features for advanced sensing applications (e.g., activity recognition).
Finally, our measurements are supplemented with marker-based motion tracking \ac{gt} data, providing, for the first time, a reliable reference to assess both \ac{md} and tracking traces.

\section{System Overview}
The system is composed of one \ac{tx} and two \ac{rx} antenna arrays exchanging standard-compliant single-carrier IEEE 802.11ay data packets. Every packet includes $12$ \acl{trn} fields,
each transmitted using a different antenna \ac{bp}, $b$, that are used to estimate the \ac{cir}.
The \ac{cir} estimates are processed to remove the \ac{to} and the \ac{fo} caused by the clocks asynchrony
and are used to track the movement of a subject and to extract its \ac{md} signature.
\begin{figure*}[hbt!]
    \centering
        \subfloat[][\label{fig:sit-gt}Sit/stand \ac{md} with \acs{gt}.]
    {\includegraphics[width=\size\textwidth]{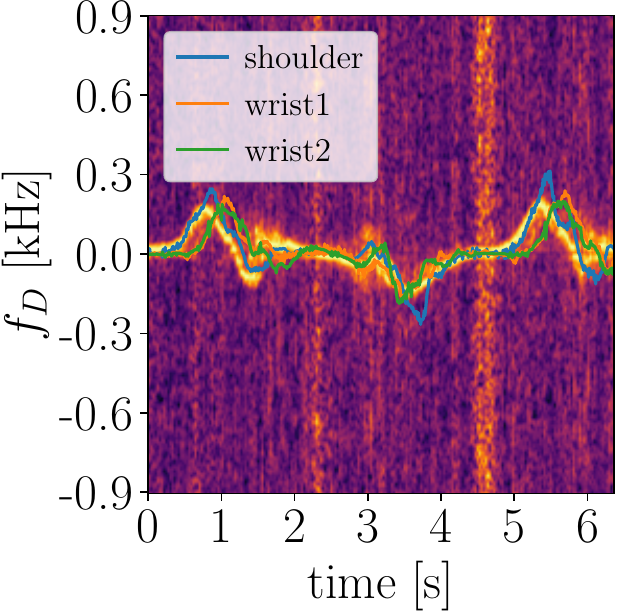}} \,
        \subfloat[][\label{fig:ex1}Walk. \ac{md} + \ac{gt} (\ac{rx}1).]
    {\includegraphics[width=\size\textwidth]{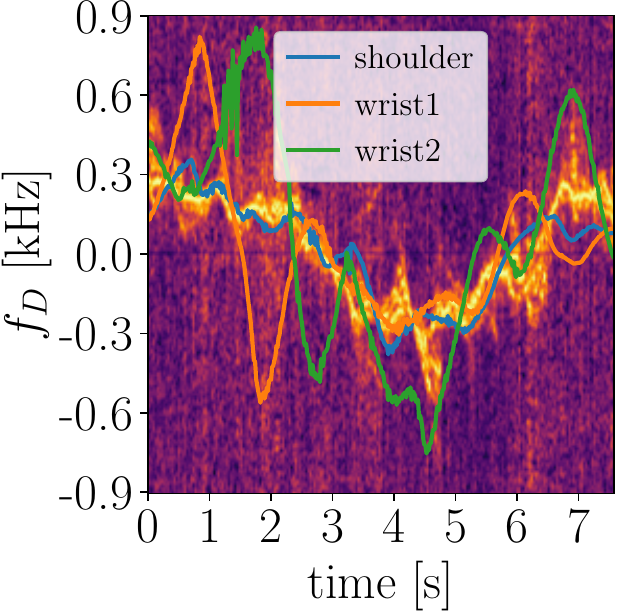}} \,
        \subfloat[][\label{fig:ex2}Walking \ac{md} (\ac{rx}2-P1).]
    {\includegraphics[width=\size\textwidth]{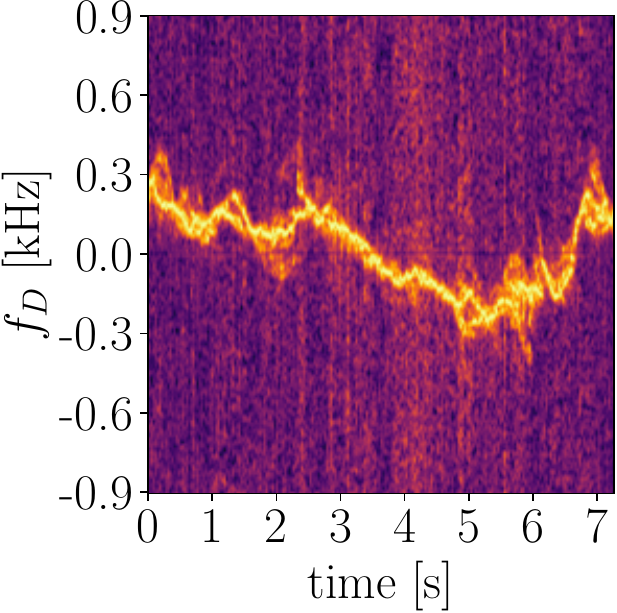}} \,
        \subfloat[][\label{fig:ex3}Walking \ac{md} (\ac{rx}2-P3).]
    {\includegraphics[width=\size\textwidth]{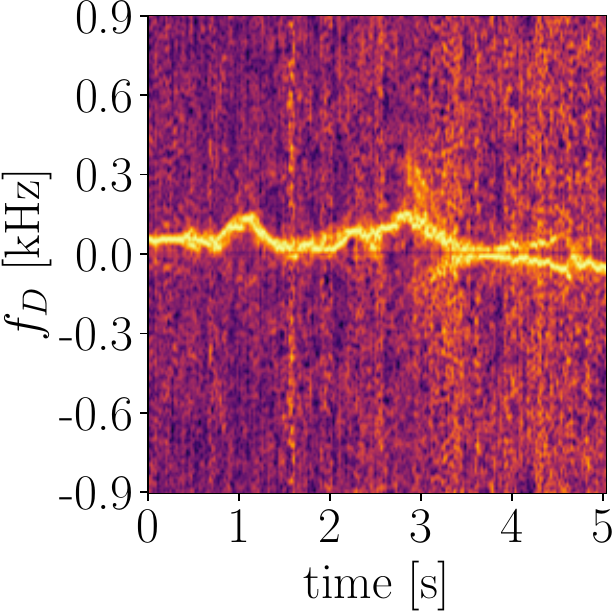}} \,
        \subfloat[][\label{fig:tracking}Example of tracking.]
    {\includegraphics[width=0.156\textwidth]
    {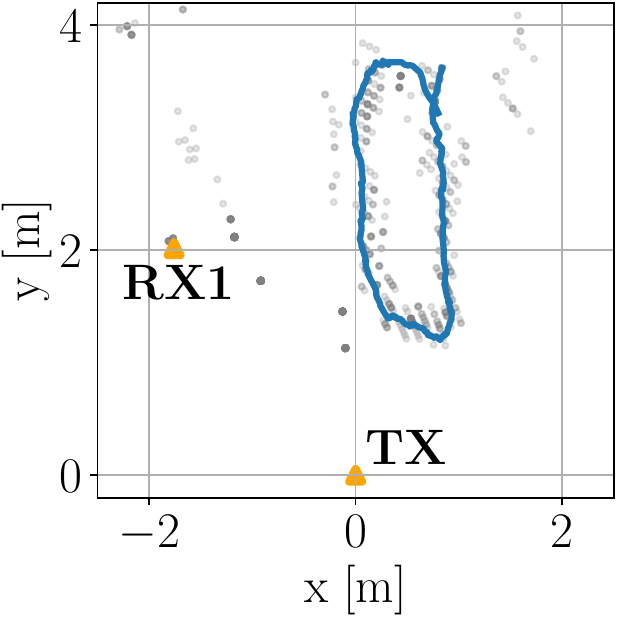}} \quad\,
        \subfloat[][\label{fig:setup}Setup scheme.]
        {\raisebox{14.5pt}        {\includegraphics[width=0.126\textwidth] {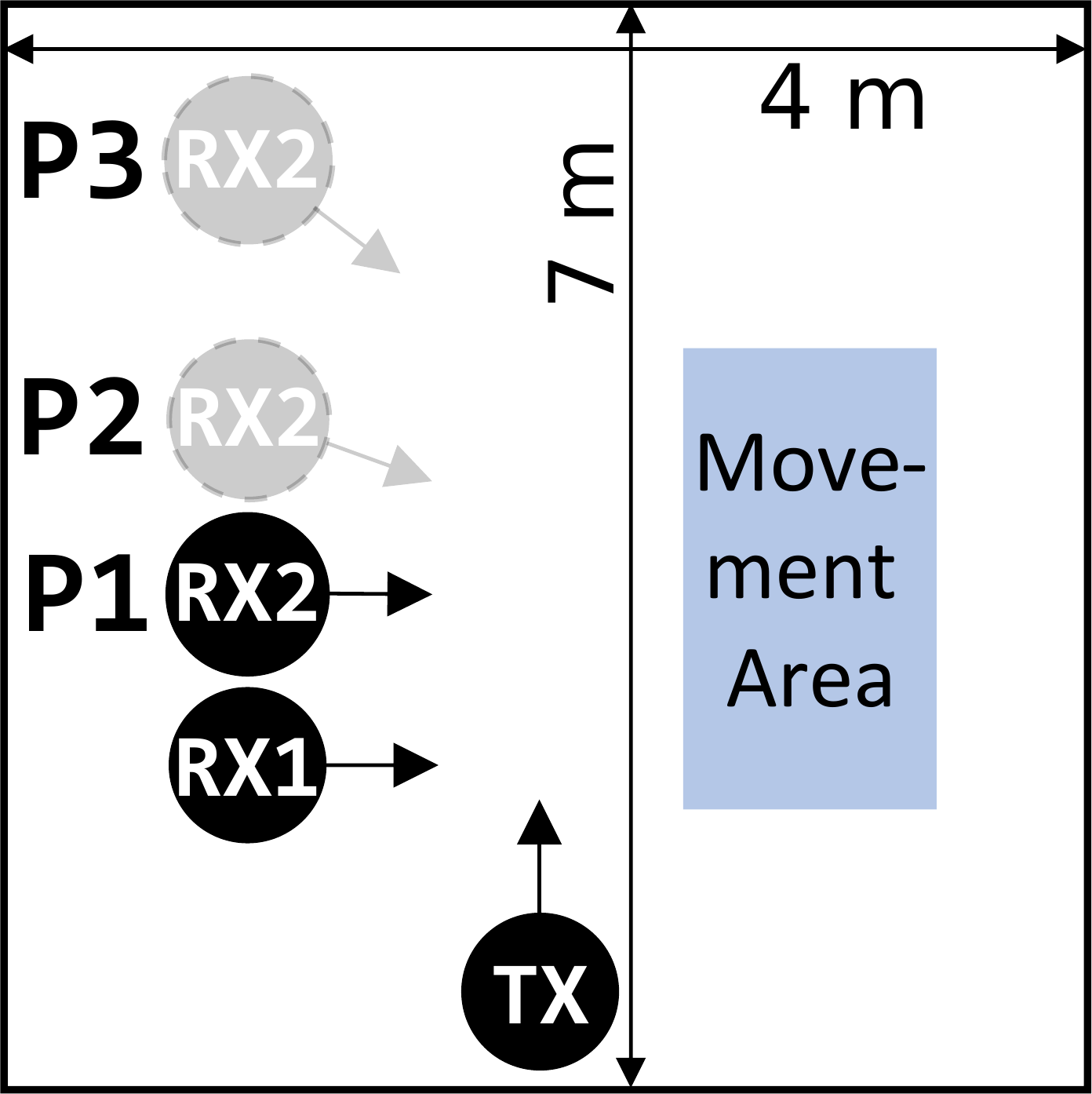}}}
    \vspace{-6pt}
    \caption{Experimental results and measurement setup scheme.}
    \label{fig:results}
    \vspace{-9pt}
\end{figure*}
\subsection{Channel Model}
At a specific time $k$, the estimated \ac{cir} is represented as a vector of $L$ complex channel gains, called \emph{taps}, and indicized by $\ell$. Each tap corresponds to a specific propagation delay (and, therefore, path length) of the transmitted signal.
Indicating with $T$ the time between two consecutive channel estimations, with $N_\ell(kT)$ the number of reflections corresponding to tap $\ell$,
and with $A_n(kT)$ the complex signal attenuation coefficient, the \ac{cir} is expressed as
    $h(k, \ell) = \sum_{n=1}^{N_{\ell}(kT)} A_{n}(kT) e^{j2\pi [f_{D, n}(kT) + f_{\rm off}(kT)]kT}$,
where $f_{D,n}(\cdot)$ and $f_{\rm off}(\cdot)$ represent, respectively, the Doppler frequency induced by targets motion in path $n$ and the \ac{fo}.
Since training fields are transmitted with different \acp{bp}, corresponding to different directions, each of them yields a different \ac{cir}, denoted by $h_b(k, \ell)$.

\subsection{Target tracking and 
\ac{md} extraction}

\shortpar{\ac{to} compensation.}
As explained in \cite{pegoraro2023jump}, it is fundamental to compensate for the TO in order to correctly localize a target. In this work, we assume that a \ac{los} path between \ac{tx} and \ac{rx} is available and we use it as a common reference to align the \ac{cir} estimates in time. 
For that, according to a dynamic threshold, we detect the peaks in the \ac{cir} magnitude and select the first one, as the \ac{los} always corresponds to the shortest path.
After that, the \ac{cir} is shifted to make the \ac{los} peak aligned with the first \ac{cir} tap. This operation is repeated for all time steps $k$ and all \acp{bp} $b$.
\\ \shortpar{Target detection.}
After this operation, to facilitate the detection of dynamic targets, static paths are removed from the \acp{cir}.
Since static paths are constant across time, we consider the time-averaged \ac{cir}, denoted as $\overline{h}_b(\ell)$,
to be a good estimate of the static background.
Then, the foreground CIR amplitude is computed as $\tilde{h}_b(k, \ell) = \max(|h_b(k, l)|-\overline{h}_b(l), 0)$. At this point, dynamic target detections at time $k$ are computed by applying a peak detection algorithm to $\sum_{b=0}^{N_b-1} (\tilde{h}_b(k, \ell))^2$, being $N_b$ the number of \acp{bp}. Finally, assuming the locations of \ac{tx} and \acp{rx} are known, exploiting bistatic geometry \cite{pegoraro2023jump} we compute the distance of the target from \ac{tx} and the path's \acl{aod}, thus, localizing the target.
\\
\shortpar{Target tracking.}
Based on the detected target locations, an \ac{ekf} is built to track the movement of the targets, assuming they approximately follow a constant velocity model.
\\
\shortpar{\ac{fo} compensation and \ac{md} extraction.}
To compute the \ac{md} spectrograms, we utilize multiple subsequent received packets.
However, since the \ac{tx} and \acp{rx} \acp{lo} are not precisely synchronized, every received packet is affected by a different \ac{fo} that spoils the \ac{md} reconstruction. 
Noting that every tap of a \ac{cir} have the same \ac{fo}, it is possible to isolate the \ac{fo} from a static path and use it to correct all paths.
In our case, we use the \ac{los} path as a reference to extract the \ac{fo}.
Let $\phi_{\rm off}(k) = 2\pi f_{\rm off}(kT)kT$ be the phase of the \ac{los} path at time $k$. Then, the \ac{fo}-corrected \ac{cir} is given by $h'_b(k, \ell)=e^{-j \phi_{\rm off}(k)}h_b(k, \ell)$.
After the correction, the resulting \ac{cir} can be used to compute the target's \ac{md} spectrogram as in \cite{pegoraro2023jump}.

\section{Experimental Results}


\shortpar{Implementation and Dataset.}
The experimental setup is a customization of MIMORPH \cite{lacruz2021a}, based on the $60$~GHz IEEE~802.11ay standard. It comprises 2 Xilinx Zynq UltraScale+ RFSoC ZCU111 boards (one for \ac{tx} and one for \acp{rx}) and $3$ Sivers EVK06002/00 antenna arrays with a radio frequency bandwidth of $1.76$~GHz.
\ac{tx} and \acp{rx} are disposed according to 3 possible different setups, as shown in \fig{fig:setup}. \ac{tx} and \ac{rx}1 are always positioned at the same locations, while \ac{rx}2 is located at P1, P2, or P3, based on the setup. \acp{rx} point towards the movement area, to maximize the received \acl{snr} of target's reflections.
We acquired data from both \acp{rx} simultaneously.
In total, we collected $72$ sequences, $24$ per setup, $36$ per array. Within each setup, the subject performed two activities: \emph{walking} along an oval trajectory or \emph{sitting down/standing up} from a chair.
Moreover, we acquired \emph{\acl{gt}} data through a marker-based motion capture system with millimeter level accuracy. Markers were placed on the antenna arrays and on some subject's body parts, namely: wrists, head, and shoulder. We use the head markers to precisely identify the location of the subject, the shoulder markers as an estimate of the chest movement, and the wrist markers as indicative of the arms movement.
\\
\shortpar{Results.}
\fig{fig:ex1}-\subref*{fig:ex3} show the \ac{md} obtained from the same walking activity when reflections are captured at different locations. We observe that the frequency of the Doppler shift decreases in P2 and P3.
Since the bistatic angle $\beta$ (the angle between \ac{tx} and \ac{rx} with vertex at the target) affects the Doppler frequency through a multiplicative factor $\xi = \cos (\beta/2)$, moving \ac{rx}2 to P2 and P3 corresponds to approaching $\beta = \pi$, and, therefore, $\xi$ decreases. Hence, our observation is in line with what expected from the theory.
\fig{fig:sit-gt}-\subref*{fig:ex1} show the \ac{md} reconstruction along with the estimate from the \ac{gt} data. In \fig{fig:sit-gt}, with sit/stand activity, body parts undergo a similar movement and this is reflected in both \ac{md} and \ac{gt} curves having the same shape. In \fig{fig:ex1}, the strongest \ac{md} component follows the movement of the chest, the largest body part, while the wrists movement align with the weaker components.
\fig{fig:tracking} shows a sample tracking result, where in blue we show the EKF output.


\section{Conclusions and outlook}
In this work, we presented the first experimental setup for \emph{asynchronous and multistatic} \ac{isac} in the \ac{mmwave} frequency band. We collected a dataset featuring $1$ \ac{tx} and $2$ \acp{rx} simultaneously receiving IEEE~802.11ay data packets. The dataset is supplemented with \ac{gt} data that allows comparing, for the first time, \ac{md} traces with a quantitative and reliable reference. 
We show that \ac{los}-based asynchrony resolution allows for accurate multistatic sensing without any kind of synchronization.
This paves the way to new research to combine the information received at different nodes.

% ...

\bibliographystyle{ACM-Reference-Format}
\bibliography{biblio}

\end{document}